\documentclass[twocolumn]{aastex62}

\usepackage{amsmath} 

\graphicspath{./}

\received{January 11, 2019}
\accepted{May 15, 2019}

\shorttitle{IC scattering of starlight in the kpc-scale jet in Cen A}
\shortauthors{Tanada et al.}

\usepackage{lineno}
\begin{document}
\title{Inverse Compton scattering of starlight in the kiloparsec-scale jet in Centaurus A: The origin of excess TeV $\gamma$-ray emission}

\correspondingauthor{Kazuhisa Tanada}
\email{arcmin\_phase19@fuji.waseda.jp}

\author{K. Tanada}
\affiliation{Research Institute for Science and Engineering, Waseda University, 3-4-1, Okubo, Shinjuku, Tokyo, 169-8555, Japan}

\author{J. Kataoka}
\affiliation{Research Institute for Science and Engineering, Waseda University, 3-4-1, Okubo, Shinjuku, Tokyo, 169-8555, Japan}

\author{Y. Inoue}
\affiliation{Interdisciplinary Theoretical and Mathematical Science Program (iTHEMS), RIKEN, Saitama 351-0198, Japan}
\affiliation{Kavli Institute for the Physics and Mathematics of the Universe (WPI), The University of Tokyo, Kashiwa, Chiba 277-8583, Japan}

\begin{abstract}

Centaurus A (Cen~A) is the nearest active radio galaxy, which has kiloparsec (kpc) scale jets and {giant lobes detected by various instruments in radio and X-ray frequency ranges}.
The $Fermi$--Large Area Telescope and High Energy Stereoscopic System (HESS) confirmed, that Cen~A is a very high-energy (VHE; $> 0.1$~TeV) $\gamma$-ray emitter with a known spectral {softening} in the energy range from a few GeV to TeV.
In this work, we consider a synchrotron self-Compton model in the nucleus for the broad band spectrum {below the break energy} and an external Compton model in kpc-scale jets for the $\gamma$-ray excess.
Our results show that the observed $\gamma$-ray excess can be suitably described by the inverse Compton scattering of the starlight photons in the kpc-scale jets, which is consistent with the recent tentative report by the HESS on the spatial extension of the TeV emission along the jets.
Considering the spectral fitting results, the excess can only be seen in Cen~A, which is probably due to two factors: (1) the host galaxy is approximately 50 times more luminous than other typical radio galaxies and (2) the core $\gamma$-ray spectrum quickly decays above a few MeV due to the low maximum electron Lorentz factor of $\gamma_{\rm c}=2.8 \times 10^3$ resulting from the large magnetic field of 3.8~G in the core.
By the comparison with other $\gamma$-ray detected radio galaxies, we found that the magnetic field strength of relativistic jets scales with the distance from the central black holes $d$ with $B (d) \propto d^{-0.88 \pm 0.14}$.

\end{abstract}


\section{Introduction} \label{sec:intro}
Centaurus A (Cen~A, also known as NGC 5128) hosts the nearest active galactic nucleus (AGN) at a redshift of $z = 0.00183$, with {a radio jet} and large-scale Fanaroff--Riley type~I (FR~I) radio morphology.
The jet of Cen~A is oriented at a large angle with respect to the line of sight of the observer, $\theta = 15$--$80^\circ$, as estimated by very-long-baseline interferometry {\citep[VLBI;][]{1998AJ....115..960T, 2003ApJ...593..169H}.}
The mass of the central supermassive black hole is estimated to be $M_{\rm BH} = (5.5 \pm 3.0) \times 10^7 \ \rm M_{\odot}$ using stellar kinematics studies \citep{2009MNRAS.394..660C}.
In the X-ray region, the extended jet structures, such as lobes and knots, are observed by $Chandra$.
{According to the long-term monitoring, the spectral indices of the majority of the knots are steep, which indicates that the emission mechanisms of them are synchrotron radiation.}
In the core region, the Seyfert-like disk radiation dominates the X-ray spectrum; however, the jet core emission also gives contribution in the range of $\sim$1--10\% to the entire core X-ray emission.
The jet X-ray component has a longer timescale of variability than the Seyfert-like disk emission \citep{2007ApJ...665..209M, 2011ApJ...743..124F}.
In the $\gamma$-ray bands, both high-energy (HE; $>$ 0.1~GeV) and very high-energy (VHE; $>$ 0.1~TeV) $\gamma$ rays have been detected in Cen~A.
The recent $Fermi$--Large Area Telescope (LAT) observation shows a spectral hardening above $\sim 2.8$~GeV \citep{2013ApJ...770L...6S, 2018A&A...619A..71H}.
In addition, this hard component is {smoothly} connected to the TeV emission detected by the High Energy Stereoscopic System (HESS) \citep{2018A&A...619A..71H}.
The spectral energy distribution (SED) below the break energy can be described by one-zone synchrotron self-Compton (SSC) models \citep{2010ApJ...719.1433A}.
For the modeling of the GeV--TeV hardness component, it was assumed that the HE and VHE emissions were produced in the core similarly to that of the two-zone SSC model \citep{2018A&A...619A..71H}, SSC combined with the dark matter particles annihilation model \citep{2017PhRvD..95f3018B}, SSC + external Compton (EC) combined with the disk photons model, and SSC + EC + photohadronic model \citep{2018MNRAS.478L...1J}.
However, recently, a new analysis technique, with the improved understanding of the point spread function (PSF) of HESS, revealed that the emission of the VHE $\gamma$-ray is extended along the kiloparsec (kpc) scale jet \citep{2018TeVPA}.
Considering this study, kpc-scale jet models explaining the TeV emission by EC with inner jet photons, such as \cite{2019MNRAS.483.1003B} have been reported.
Furthermore, \cite{2011MNRAS.415..133H} suggest that the EC radiation at the kpc-scale jets represents the GeV--TeV $\gamma$-ray emission.
{In addition, there is a model with the isotropic kpc-scale pair halo formed due to the absorption of the VHE gamma-ray emission of the core on the starlight \citep{2006MNRAS.371.1705S}.}

In this work, we aim to explain the overall SED of Cen~A using a leptonic model involving the SSC model + EC combined with the starlight model considering the core and kpc-scale jets.
The double-peaked SED was fitted with the one-zone SSC model in the core region up to a few GeV and the GeV--TeV hardness component was fitted with the Compton scattering of starlight photons due to the ultrarelativistic electrons in the jet knots.
The X-ray observations of the knots of Cen~A with $Chandra$ are presented in Section~\ref{sec:chandra}.
In Section~\ref{sec:results}, we investigate the energy densities of various seed photons to determine the dominant photons for inverse Compton (IC) scattering at the kpc scale.
Then, the results of the fitting using the SSC method and EC combined with the starlight model are presented.
In Section~\ref{sec:discussion}, we explain the detection of the spectral hardening in Cen~A by discussing the relationship between the magnetic field strength of the emission zones and the distance from the core. Our conclusions are presented in Section~\ref{sec:conclusion}.
In this paper, we assume standard cosmology with $H_0=70 \ \rm km \ s^{-1} \ Mpc^{-1}$ and $\Omega_m=0.29$ \citep{2014ApJ...794..135B}. This corresponds to a linear scale of 1~arcsec$ \ =\ $18~pc at a luminosity distance of $D_L$ =3.8~Mpc. The errors in this work are at the 1-$\sigma$ confidence level unless stated otherwise.

\section{$Chandra$ Observation} \label{sec:chandra}
The Advanced CCD Imaging Spectrometer (ACIS-I) detector on board the $Chandra$ X-ray Observatory has an angular on-axis resolution of $\sim 0.5 \arcsec$ and it operates in the 0.2--10~keV range.
Its high angular resolution enables the investigation of the nonthermal emission from the knots in the kpc-scale jet of Cen~A.
Although more than 30 X-ray knots have been observed \citep{2002ApJ...569...54K, 2010ApJ...708..675G}, 
approximately half of them have steep spectra, indicating the high-energy part of the synchrotron emission, and it is unlikely that they are related to the observed gamma-ray excess. 
In this study, we consider the two brightest hard spectral ($\Gamma_{\rm X} < 2$) knots, AX2 and BX2. 
In addition to them, for comparison, we also consider two bright soft spectral knots, AX1A and AX1C. 
These four knots were also detected by radio observations at 4.8, 8.4, and 22~GHz by the Very Large Array (VLA) operated by the National Radio Astronomy Observatory (NRAO) \citep{2002ApJ...569...54K, 2007ApJ...670L..81H, 2010ApJ...708..675G}.
We selected three observations (ObsID 7800, 8489, 8490; PI Kraft), with exposure times of 90.8 (on April 17, 2007), 93.9 (on May 8, 2007) and 94.4~ksec (on May 30, 2007), respectively.
{According to \cite{2010ApJ...708..675G}, there is no flux variability for AX1A and AX2, and there are small variabilities (less than $\simeq 20 \%$) for AX1C and BX2, during the time range.}
The data were analyzed using the $Chandra$ Interactive Analysis of Observations (CIAO) software v4.8 and the $Chandra$ Calibration Database (CALDB) v4.7.4. The spectral analysis was performed using the X-Ray Spectral Fitting Package (XSPEC) v12.9.

We extracted the 0.5--9~keV spectrum of each knot using a region with a radius of $3\arcsec$ and analyzed the local background in an annulus around the source with an inner radius of $3\arcsec$ and an outer radius of $6\arcsec$. 
Those non-interesting bright regions (detected by the CIAO tool {\tt wavdetect} with $> 3 \ \sigma$ significance) that are included in the region of interest, are omitted.
The spectral fitting results of the four knots are shown in Table~\ref{table:chandra_knots}.
We use the same X-ray knot IDs as those in Figure~2 in \cite{2010ApJ...708..675G}.
For the fitting of the merged spectra, we adopted the model $\tt phabs \times powerlaw$ in the XSPEC, where $\tt phabs$ corresponds to the absorptions in the Galaxy and Cen~A and $\tt powerlaw$ is the nonthermal power-law emission from the knots.

In addition, we estimated the width of the knot perpendicular to the direction of the propagation from the $Chandra$ X-ray image in the 0.5--10~keV bandpass by
\begin{equation}
	\sigma_{\rm Knot} = \sqrt{{\sigma_{\rm obs}}^2 - {\sigma_{\rm PSF}}^2},
	\label{eq:sigma_knot}
\end{equation}
where $\sigma_{\rm Knot}$ is the full width at maximum height (FWHM) width of the knot, $\sigma_{\rm obs}$ is the observed width (FWHM) of the knot, and $\sigma_{\rm PSF}$ is the width of the PSF (FWHM).
{For this analysis, we used only one observation (OBSID: 8489) to avoid uncertainties that would be introduced by combining the PSFs of two or more observations at different roll angles and off-axis positions.}
The obtained width of the knots and their distance from the core are shown in Table~\ref{table:chandra_knots}.

\begin{deluxetable*}{cccccccc}[htb]
\tablecaption{Spectral fitting results of the jet knots of Cen~A}
\tablecolumns{7}
\tablewidth{0pt}
\tablehead{
\colhead{Knot} &
\colhead{${N_{\rm{H}}\tablenotemark{a}}$} &
\colhead{${N_0}\tablenotemark{b}$} &
\colhead{$\Gamma_{\rm X}\tablenotemark{c}$} &
\colhead{Reduced $\chi ^2$ ($\chi ^2/$dof)} &
\colhead{Distance\tablenotemark{d}} &
\colhead{${\sigma_{\rm Knot}}\tablenotemark{e}$}
}
\startdata
AX1A & $0.54 \pm 0.03$ & $7.50 \pm 0.52$ & $2.25 \pm 0.06$ & $0.94 \ (143.0/152)$ & 0.9 & $0.8 \pm 0.4$\\
AX1C & $0.59 \pm 0.03$ & $11.9 \pm 0.1$ & $2.19 \pm 0.07$ & $1.08 \ (194.1/179)$ & 0.9 & $1.7 \pm 0.4$\\ 
AX2 & $0.51 \pm 0.09$ & $1.76 \pm 0.28$ & $1.93 \pm 0.14$ & $1.03 \ (63.0/61)$ & 1.1 & $2.7 \pm 0.7$\\
BX2 & $0.17 \pm 0.03$ & $3.97 \pm 0.25$ & $1.74 \pm 0.06$ & $1.00 \ (163.4/163)$ & 3.5 & $3.5 \pm 0.2$\\
\enddata
\tablenotetext{a}{Hydrogen column density from the Galaxy and Cen~A in units of $\times 10 ^{22} \ \rm cm^{-2}$.}
\tablenotetext{b, c}{\ \ \ \ Photon index and normalization at 1~keV of the power-law model. The normalization is in units of $\times 10^{-5} \ \rm {ph~cm^{-2} \ s^{-1} \ keV ^{-1}}$.}
\tablenotetext{d}{Distance from the nucleus in units of $\times 10^{21} \ \rm cm$. These values are obtained from \cite{2002ApJ...569...54K}.}
\tablenotetext{e}{Size of the source region in units of $\times 10^{19} \ \rm cm$, measured using Eq.~\ref{eq:sigma_knot}.}
\label{table:chandra_knots}
\end{deluxetable*}

\section{Results}\label{sec:results}
\subsection{Estimation of the Photon Energy Density Along the Jet Axis} \label{sec:phE_density}
\begin{figure}[ht!]
  \centering
  \includegraphics[scale=0.45]{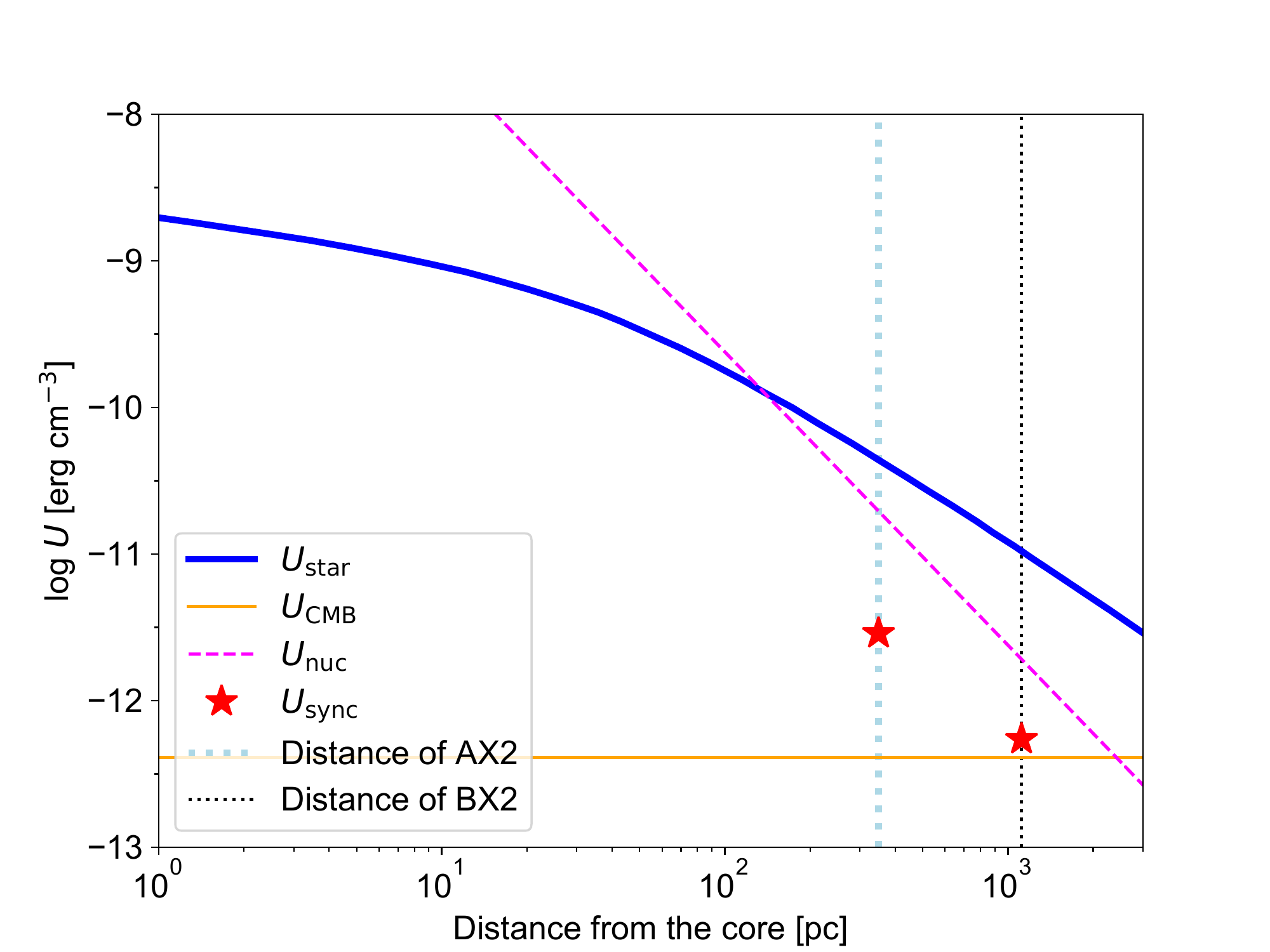}
  \caption{Energy densities of various seed photons along the jet axis. The blue solid line represents the starlight photon energy density ($U_{\rm star}$). The energy density of the CMB ($U_{\rm CMB}$) and the synchrotron emission from the nucleus jet ($U_{\rm nuc}$) are represented by the orange solid line and magenta dashed line, respectively. The synchrotron emissions of the AX2 and BX2 knots ($U_{\rm sync}$) are represented by red stars. The light blue and black dotted lines represent the distances of the AX2 and BX2 knots from the nucleus, respectively.}
  \label{fig:u_photon}
\end{figure}

To determine the dominantly scattered seed photons by the electrons in the kpc-scale jet, we analyzed the energy densities of the ambient radiation fields (such as the synchrotron photons from the nucleus, synchrotron photons in the knots, starlight photons from the host galaxy, and the cosmic microwave background (CMB)) along the jet axis, shown in Figure~\ref{fig:u_photon}, measured 
in the rest frame of the outer jet at different distances from the core.
The emissivity profile of the starlight $j_{\rm star}$ is expected to follow the galactic mass distribution; thus, we can obtain the energy density profile for the starlight emission using an approximate calculation by integrating $j_{\rm star}$ along a ray and the solid angle, as $U_{\rm star} = (1/c)\int j_{\rm star}(r) {\rm d}s \ {\rm d}\Omega$ \citep{2006MNRAS.370..981S}.
In this study, we used the result from the detailed calculation of the starlight emission profile for Cen~A reported by \cite{2006MNRAS.371.1705S}.
For the analysis of the energy density of the synchrotron photons from the nucleus in the kpc-scale jet rest frame, we assume that the radiation enters the considered emitting region directly from the jet base.
Then, as discussed in \cite{2003ApJ...597..186S}, the energy density can be represented by 
\begin{equation}
		U_{\rm nuc} = \frac{L_{\rm nuc}}{4 \pi r^2 c} \left( \frac{2 \Gamma_{\rm nuc}}{\delta_{\rm nuc}} \right) ^3 \frac{1}{(2 \Gamma)^2}
  		\label{eq:u_nuc}
\end{equation}
where $L_{\rm nuc}$ is the synchrotron luminosity of the nuclear jet, $\delta_{\rm nuc}$ is the nuclear Doppler factor, which reflects a jet viewing angle of $\theta = 30 ^\circ$ and nuclear jet bulk Lorentz factor of $\Gamma_{\rm nuc} = 1.0$, and $\Gamma = 1.0$ is the bulk Lorentz factor of the emission region.
{The bulk Lorentz factors of the nuclear and kpc-scale jet are derived from the VLBI and VLA observations \citep{1998AJ....115..960T, 2010ApJ...708..675G}.}
It should be noted, that $L_{\rm nuc}$ was set to $1.7 \times 10^{42} \ \rm erg \ s^{-1}$, which is the expected luminosity at the synchrotron peak frequency of $\simeq 6.3 \times 10^{13}$~Hz for the SSC model, which is discussed in a later section.
We also analyzed the energy density of the CMB, as seen in the comoving frame, by
\begin{equation}
		U_{\rm CMB} = a T_{0}^4 \Gamma^2 (1+z)^4,
  		\label{eq:u_CMB}
\end{equation}
where $a = 7.56 \times 10^{-15} \ {\rm erg \ cm^{-3} \ K^{-4}}$ and $T_0 = 2.7$~K is the current temperature of the CMB, at $z = 0$ {\citep{2015APh....63...55A}}. In this case, we calculated it at the bulk Lorentz factor of $\Gamma = 1.0$ and $z = 0.00183$, while the obtained energy density had a constant value of $U_{\rm CMB} = 0.26 \ \rm eV \ cm^{-3}$.
Finally, the energy density of the synchrotron photons within the jet knots was represented by $U_{\rm sync} = (4 \pi / c) \int I_{\nu, \rm sync} \ {\rm d}\nu$, where $I_{\nu, \rm sync}$ is the synchrotron intensity \citep{1996ApJ...463..555I}.
For a possible spectral connection between the X-ray and GeV--TeV component,
we consider the AX2 and BX2 knots in this section, with energy densities of $U_{\rm sync} = 2.9 \times 10^{-12} \ \rm erg \ cm^{-3}$ and $U_{\rm sync} = 5.5 \times 10^{-13} \ \rm erg \ cm^{-3}$, respectively, obtained from the EC models in Section~\ref{sec:knots_emission}.

Figure~\ref{fig:u_photon} shows the energy density profiles for various IC seed photons of the above analysis, and it indicates that the starlight photon fields have higher energy densities than the others at more than a few hundred pc.
Thus, the starlight is a strong candidate as IC seed photons for the kpc-scale jet knots, such as AX2 and BX2 located at a distance of 350~pc and 1100~pc from the nucleus, respectively. 
It should be noted, that in the case of M87 (an extensively studied FR~I radio galaxy) the starlight component is weaker than that in Cen~A \citep{2011MNRAS.415..133H}.
In addition, the total K-band luminosity of the galaxy obtained from the Two Micron All Sky Survey (2MASS) of Cen~A is approximately 50 times greater than that of other typical radio-detected AGNs \citep{2006A&A...447...97B}.
{We note that, if we assume a bit larger Lorentz factor ($\Gamma_{nuc} > 3$) for the sub-parsec scale jet, contribution from the nuclear jet emission may dominate the photon density than anticipated starlight emission at the kpc scale (see, Eq.~\ref{eq:u_nuc}). 
However, this is an extreme case in which kpc-scale jet is situated exactly at the downstream of a starlight jet having a fixed opening angle, which seems unrealistic
as implied from the $Chandra$ X-ray images \citep{2002ApJ...569...54K}.
}

\subsection{SED and Modeling} \label{sec:gev_hardness}
The emission from the core of Cen~A is represented by the double-peaked shape observed in many blazars.
Therefore, we attempted to model the overall SED of the core up to a few GeV with a standard homogeneous one-zone SSC model developed and widely used for BL~Lac objects \citep[for details see][]{1996ApJ...463..555I,1999ApJ...514..138K,2003ApJ...593..667M, 2016ApJ...828...13I}.
The recent $Fermi$-LAT analysis provided evidence for the spectral hardening from the photon index of $2.70 \pm 0.02$ to $2.31 \pm 0.07$ at a break energy of $\simeq 2.8$~GeV, at a level of $4 \ \sigma$ \citep{2018A&A...619A..71H}.
{This hardness component, including the HESS spectrum above the break energy, exceeds over the power-law extrapolation of the gamma-ray spectrum obtained by likelihood analysis below the break energy.}
Then, we assumed that the high-energy $\gamma$-ray emission above a few GeV resulted from the IC scattering of the starlight photons by the ultrarelativistic electrons in the kpc-scale jets.
The SSC and EC models consider an electron energy distribution in the form of a broken power-law as
\begin{equation}
N(\gamma) = 
  \begin{cases}
  	K (\gamma / \gamma_{\rm brk})^{-p_1}  & \text{($\gamma \leq \gamma_{\rm brk}$)} \\
	K (\gamma / \gamma_{\rm brk})^{-p_2}  & \text{($\gamma > \gamma_{\rm brk}$)},
  \end{cases}
\end{equation}
for an electron Lorentz factor $\gamma$ ($\gamma_{\rm min} < \gamma < \gamma_{\rm c}$), where $\gamma_{\rm brk}$ represents the electron break Lorentz factor, at which the radiative cooling time equals the dynamic timescale. 
{Above the electron maximum Lorentz factor $\gamma_{\rm c}$, the electron distribution follows an exponentially cutoff function.}
The parameter $p_1$ represents the low-energy electron index between $\gamma_{\rm min}$ and $\gamma_{\rm brk}$ and $p_2$ represents the high-energy electron index between $\gamma_{\rm brk}$ and $\gamma_{\rm c}$. The electron density is represented by $K$.
Further physical parameters of this model are the source radius, $R$, the magnetic field, $B$, the Doppler factor, $\delta = 1/[\Gamma (1-\beta \cos \theta)]$, where $\beta$ is the bulk speed of the plasma moving along the jet, the bulk Lorentz factor, $\Gamma = [1-\beta^2]^{-1/2}$, and $\theta$, which is the angle between the jet axis and the line of sight. In this work, for the SED model fitting, a jet viewing angle of $\theta = 30^\circ$ was selected.
To characterize the thermal radiation from the host galaxy, a diluted blackbody was chosen at a single temperature, $T_{\rm ext, optical}$, with an optical luminosity, $L_{\rm ext, optical}$ \citep{1996ApJ...463..555I}.

In this study, the $\gamma$-ray attenuation due to pair creation was ignored. 
The $\gamma$-ray emission in the core does not extend beyond the MeV-band (See Section~\ref{sec:core_emission}).
The internal $\gamma$--$\gamma$ attenuation in the knots by optical photons only exists at a level of $\tau \sim10^{-4}$ and the extragalactic background light opacity becomes $\gtrsim 1$ only above 15~TeV \cite[e.g.,][]{2013ApJ...768..197I}.

\subsubsection{SSC Model for the Core Emission} \label{sec:core_emission}
According to the observed variability (flux doubling) timescale of the VLBI jet core emission \citep{1997ApJ...475L..93K, 2006PASJ...58..211H}, we can calculate an upper limit of the size of the core, with $t_{\rm var}$ of approximately a few days, as $R < c t_{\rm var} \delta \simeq 10^{16}$~cm for the expected $\delta \simeq 1$.
It should also be noted that monitoring of Cen~A by the VLBI shows that the sub-parsec scale jet has components with a slow apparent motion of $\sim$0.1c away from the nucleus, which implies $\Gamma = 1.005$ \citep{1998AJ....115..960T}.

Figure~\ref{fig:SED} shows the non-contemporaneous multi-wavelength $\nu F_{\nu}$ SED of Cen~A obtained from the radio to very high energy $\gamma$-ray data.
The left and right panels represent the SSC + EC model with AX2 and BX2 knots, respectively.
In the radio to optical/ultraviolet (UV) band, the archival NASA/IPAC Extragalactic Database (NED) data were used for the core emission.
The X-ray observations of $Suzaku$ and the International Gamma-Ray Astrophysics Laboratory ($INTEGRAL$) show that the X-ray emission from the core of Cen~A is dominated by Seyfert-like emission \citep{2007ApJ...665..209M, 2011A&A...531A..70B}.
To consider the pure jet emission in the X-ray band for the SSC modeling, for the jet component we used the parameters obtained from fitting the spectrum of the $Suzaku$ observation in 2009 with double power-law functions (thermal Comptonization of disk photons and jet model) previously reported by \cite{2011ApJ...743..124F}. The power-law photon index of the jet component is $\Gamma_{\rm X} = 1.6$ in the X-ray energy range of 0.5--300~keV and it is consistent with the expected slope in the radio to optical band.
The $\gamma$-ray data were obtained from the observations by the Imaging Compton Telescope (COMPTEL) of the Compton Gamma-Ray Observatory ($CGRO$) \citep{1998A&A...330...97S, 2001AIPC..587..353S} from 1991 to 1995, the $Fermi$-LAT observations from 2008 to 2016, and the HESS observations from 2004 to 2010 \citep{2018A&A...619A..71H}.
{It is reported that no significant variation is detected in the GeV--TeV energy band \citep{2017PhRvD..95f3018B, 2018A&A...619A..71H}.}

In previous studies, a large difference between low- and high-energy electron spectrum indices has been reported (e.g. $p_1 = 1.8$, $p_2 = 4.3$, see \cite{2010ApJ...719.1433A, 2018A&A...619A..71H, 2018MNRAS.478L...1J}), {which is true only if the magnetic field within the emission region is highly non-uniform.
In this study, we assume uniform magnetic field, therefore, the electron distribution breaks in its index by one power above the break energy $\gamma_{\rm brk}$, which is indicated from the steady solution of the simplified kinetic equation for electrons \citep{1996ApJ...463..555I, 2011hea..book.....L}.}
Thus, we set the difference to $\simeq 1$, and then, we obtained low- and high-energy electron spectrum indices of $p_1 = 1.90$ and $p_2 = 2.80$, respectively.
Due to this effect, we obtained a maximum electron Lorentz factor of $\gamma_{\rm c} = 2.8 \times 10^3$, which is noticeably smaller than the previously reported values of $\gamma_{\rm c} \gtrsim 10^5$.
In addition, the magnetic field of $B = 3.8$ G is significantly larger than the typical range of $B$, which is approximately several tens of mG, found for non-Blazar FR~I AGNs \citep{2003ApJ...597..166C, 2009ApJ...707...55A, 2018ApJ...860...74T, 2018Ap.....61..160B}.
As shown in Figure~\ref{fig:SED}, the overall trend of the core SED is adequately represented by the one-zone SSC model.
Further physical parameters for the core emission are the electron density of $K = 2.0 \times 10^5 \ \rm cm^{-3}$, the minimum electron Lorentz factor of $\gamma_{\rm min} = 1.0$, the electron break Lorentz factor of $\gamma_{\rm brk} = 1.8 \times 10^3$, the source radius of $R = 4.8 \times 10^{15}$~cm, the jet viewing angle of $\theta = 30 ^{\circ}$, and the Doppler factor of $\delta = 1.0$, as summarized in Table~\ref{table:SSCparam}.
{The obtained total luminosity of the core emission is $L_{\rm rad} = 1.5 \times 10^{43} \ {\rm erg~s^{-1}}$, which is within the same order of the electron kinetic jet luminosity of $L_{\rm kin} = 1.1 \times 10^{43} \ {\rm erg~s^{-1}}$.
}

\subsubsection{EC Model for the Kpc-scale Jet Emission} \label{sec:knots_emission}
In this section, the fitting of the emission from the kpc-scale jet knots, including the radio, X-ray, and GeV--TeV SED with the EC/starlight model is discussed.
In the radio band, the VLA radio observations of the knots in the Cen~A jet at 4.8, 8.4, and 22 GHz were used \citep{2010ApJ...708..675G}.
For the host galaxy emission as seed photons, the data used were obtained from the Cerro Tololo Inter-American Observatory (CTIO) \citep{1971ApJ...170L..15B}, the Orbiting Astronomical Observatory II ($OAO$-$2$) \citep{1979ApJ...228..419W}, and the 2MASS \citep{2003AJ....125..525J}.
In the X-ray band we analyzed the $Chandra$ data (used in this work) in 2007, as shown in Table~\ref{table:chandra_knots}.

\begin{figure*}[]
 \begin{minipage}[b]{0.5\linewidth}
  \centering
  \includegraphics[scale=0.4]{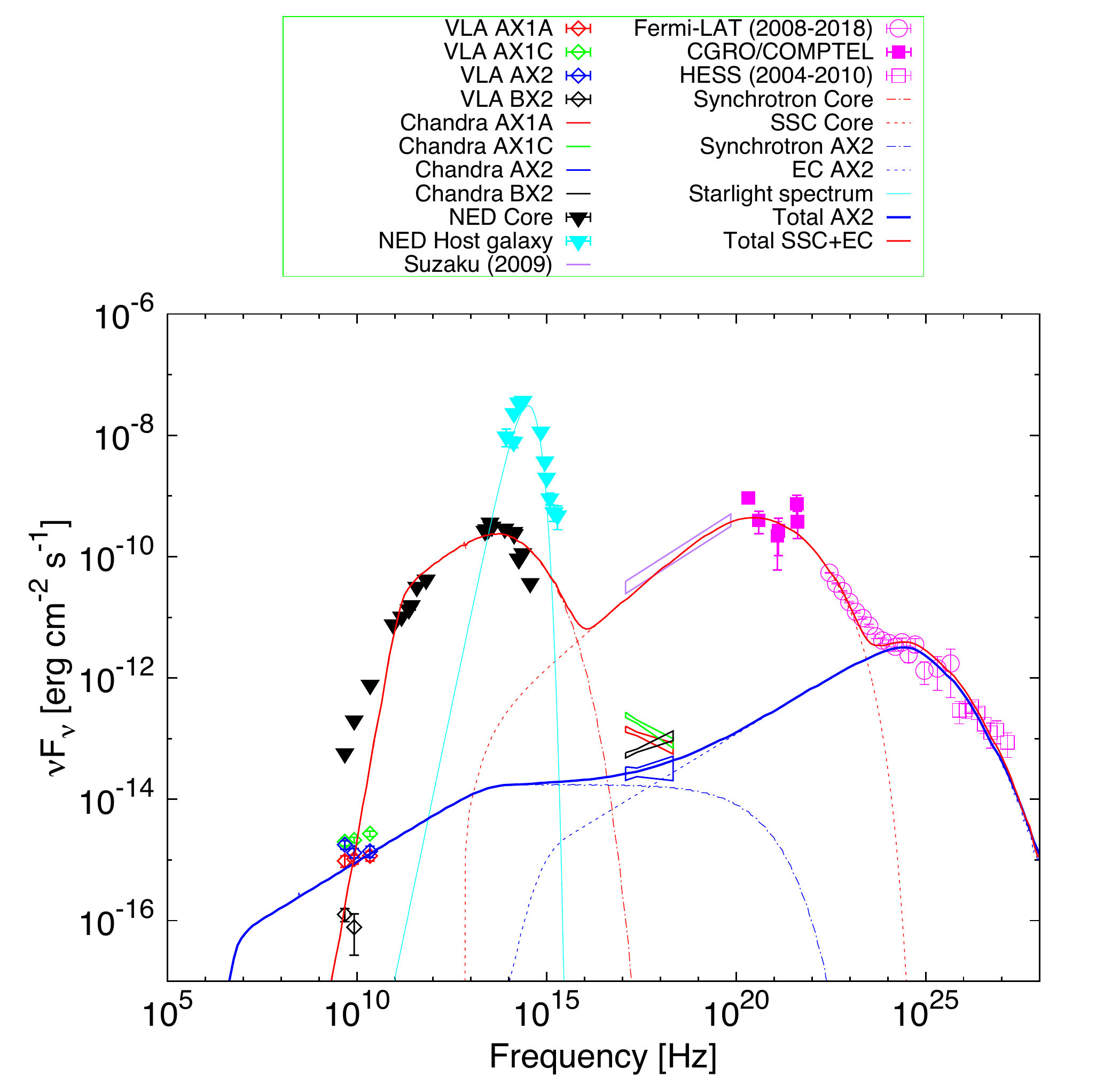}
  \label{subfig:AX2}
 \end{minipage}
 \begin{minipage}[b]{0.5\linewidth}
  \centering
  \includegraphics[scale=0.4]{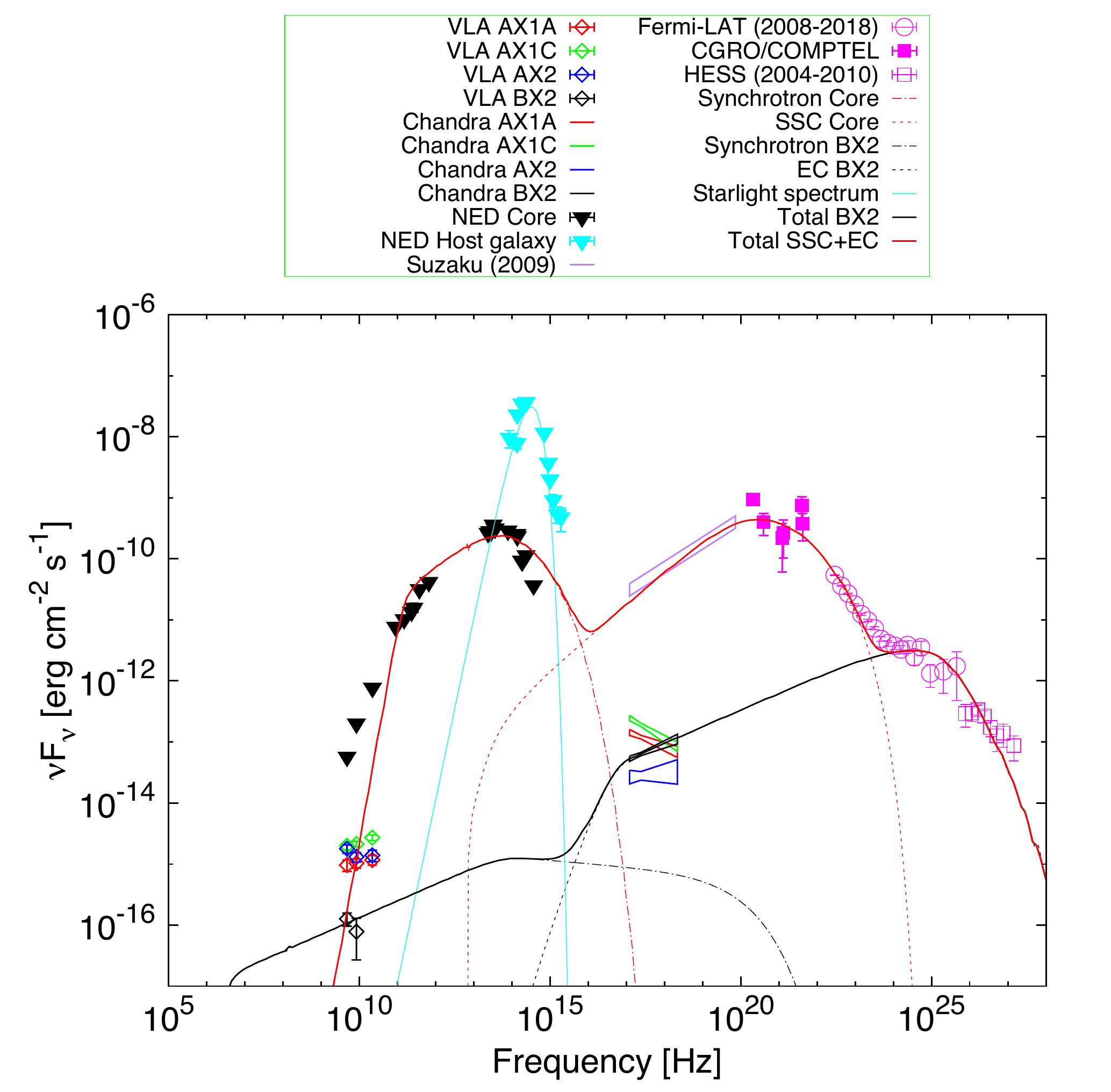}
  \label{subfig:BX2}
 \end{minipage}
\caption{Overall SED of Cen~A obtained from multi-wavelength data, using NED, $Suzaku$ bow-tie \citep{2011ApJ...743..124F}, $CGRO$-COMPTEL \citep{1998A&A...330...97S, 2001AIPC..587..353S}, $Fermi$-LAT, HESS \citep{2018A&A...619A..71H}, VLA \citep{2002ApJ...569...54K, 2007ApJ...670L..81H, 2010ApJ...708..675G}, and $Chandra$ (used in this work).
{\bf Left panel:} SSC model fitting for the core emission, represented by the red solid line, and the EC/starlight model for the kpc-scale jet emission with the AX2 knot, represented by the blue solid line.
The blue diamonds and the blue bow-tie represent the radio and X-ray spectra of AX2, respectively.
The starlight emission from the host galaxy of Cen~A is represented by the light blue solid line.
{\bf Right panel:} EC/starlight model for the kpc-scale jet emission with the BX2 knot, represented by black solid line. The core emission is the same as that in the left panel.
The black diamonds and the black bow-tie represent the radio and X-ray spectra of AX2, respectively.
}
 \label{fig:SED}
\end{figure*}

As can be seen in Figure~\ref{fig:SED}, the X-ray spectra of the AX1A and AX1C knots cannot be related to the high-energy $\gamma$-ray spectra of $Fermi$-LAT and HESS, because their photon indices are too steep ($\Gamma_{\rm X} > 2$).
Therefore, we used only the AX2 and BX2 knots for the EC modeling, that is, we fitted the excess GeV--TeV hardness component with the model of IC scattering of starlight photons in the AX2 knot (left panel) or the BX2 knot (right panel). The parameter values of the best-fit are shown in Table~\ref{table:chandra_knots}.
In both the AX2 and the BX2 cases, the overall SED of Cen~A can be fitted appropriately with the SSC + EC/starlight model, as shown in both panels of Figure~\ref{fig:SED}.
For the fitting of the SED, the source sizes of AX2 and BX2 were fixed at $R = 3.3 \times 10^{19}$~cm and $R = 3.7 \times 10^{19}$~cm, respectively, which are given in Section~\ref{sec:chandra}. 
The bulk Lorentz factors were also set as $\Gamma = 1.013$ for AX2 and $\Gamma = 1.001$ for BX2, which were estimated from the proper motional speeds of the knots monitored by the VLA \citep{2010ApJ...708..675G}.
The obtained magnetic field of $8.5 \times 10^{-4}$~G for AX2 is apparently different from that of $1.6 \times 10^{-4}$~G for BX2, because the observed radio flux of AX2 is approximately 10~times larger than that of BX2. 
Furthermore, the large difference in the electron densities of the two knots is mainly due to the difference in the Doppler factor and the X-ray flux between the AX2 and BX2 knots.
In the TeV-band, the observed photons originate from the scattering in the Klein--Nishina (KN) regime because the energy of the seed photon in the rest frame of the relativistic electron is larger than 511~keV.
This requires relatively large maximum electron Lorentz factors of $\gamma_{\rm c} \simeq 10^8$ and a hard high-energy electron spectrum slope of $p_2 \simeq 3$ for both knot models.
These results suggest that the SSC + EC/starlight model can explain the overall SED of Cen~A appropriately.
Moreover, the recently improved understanding of the PSF of HESS by new simulations and analysis techniques revealed that the VHE $\gamma$-ray of Cen~A is produced in the kpc-scale jet \citep{2018TeVPA}, which is consistent with our results.


%
%
\begin{deluxetable*}{ccccc}[htb]
\tablecaption{Fitted physical parameters for the SSC model shown in Figure~\ref{fig:SED}}
\tablecolumns{4}
\tablewidth{0pt}
\tablehead{
\colhead{Parameter} &
\colhead{Core} &
\colhead{AX2} &
\colhead{BX2} 
}
\startdata
	$K$ $[\rm cm^{-3}]$ & $2.0 \times 10^5$ & $6.5 \times 10^{-5}$ & $2.1 \times 10^{-3}$ \\
	$p_1$ & 1.90 & 2.24 & 2.45\\
	$p_2$ & 2.80 & 3.00 & 3.10	\\
	$\gamma_{\rm min}$ & 1.0 & 1.0 & 8.0 \\
	$\gamma_{\rm brk}$ & $1.8 \times 10^3$	& $8.0 \times 10^4$ & $3.0 \times 10^{5}$ \\
	$\gamma_{\rm c}$ & $2.8 \times 10^3$ & $2.0 \times 10^8$ & $4.0 \times 10^8$	\\ \hline
	$R$ [cm] &	$4.8 \times 10^{15}$ & $3.3 \times 10^{19}$ & $3.7 \times 10^{19}$ \\
	$B$ [G] & 3.8 & $8.5 \times 10^{-4}$ & $1.6 \times 10^{-4}$ \\
	$\delta$ &	1.0 & 1.1 & 1.0	\\
	$\Gamma$ &	1.005 & 1.013 & 1.001 \\
	$\theta$ $[^{\circ}]$ & 30 & 30 & 30 \\ \hline
	$T_{\rm ext, optical}$ [K] & - & $4.0 \times 10^3$ & $4.0 \times 10^3$ \\
	$L_{\rm ext, optical}$ $[\rm erg \ s^{-1}]$ & - & $3.0 \times 10^{44}$ & $3.0 \times 10^{44}$ \\ \hline
	$u_{\rm e} \ [\rm erg \ cm^{-3}]$ & 1.9 & $1.4 \times 10^{-9}$ & $2.0 \times 10^{-9}$ \\
	$u_B \ [\rm erg \ cm^{-3}]$ & $5.7 \times 10^{-1}$ & $2.9 \times 10^{-8}$ & $1.0 \times 10^{-9}$ \\
\enddata
\tablecomments{Obtained physical parameters for the emission from the core and AX2 and BX2. The parameters are the electron density, $K$, the low-energy electron spectrum slope, $p_1$, the high-energy electron spectrum slope, $p_2$, the minimum electron Lorentz factor, $\gamma_{\rm min}$, the break Lorentz factor, $\gamma_{\rm brk}$, the maximum electron Lorentz factor, $\gamma_{\rm c}$, the source radius, $R$, the magnetic field, $B$, the Doppler factor, $\delta$, the bulk Lorentz factor, $\Gamma$, the angle between the jet axis and the line of sight, $\theta$, the starlight temperature, $T_{\rm ext, optical}$, the optical luminosity of the starlight, $L_{\rm ext, optical}$, the electron energy density, $u_{\rm e}$, and the magnetic energy density, $u_B$.
}
\label{table:SSCparam}
\end{deluxetable*}
%

\section{Discussion}\label{sec:discussion}
One of the possible reasons for the detection of the GeV hardness in the SED of Cen~A is the very steep spectrum of the core emission above a few MeV.
This results from the small maximum electron Lorentz factor of $\gamma_{\rm c} = 2.8 \times 10^3$, where the other observed values of FR~I are $\gamma_{\rm c} = 10^5$ for the core of NGC~1275 \citep{2018ApJ...860...74T}, $\gamma_{\rm c} = 10^7$ for the core of M87 \citep{2009ApJ...707...55A}, and $\gamma_{\rm c} = 10^8$ for the core of Centaurus~B (Cen~B) \citep{2018CenB}.
The maximum electron Lorentz factor providing a balance between the particle acceleration and radiation cooling is represented by $\gamma_{\rm c} \varpropto \left[B/(u_B + u_{\rm soft}) \xi \right]^{1/2}$, where $u_{\rm soft}$ is the soft photon density, which is the sum of the synchrotron and the external photon contributions and $\xi$ is the efficiency of scattering for the particle acceleration, known as the gyrofactor \citep{1996ApJ...463..555I}.
According to their relationship, a higher magnetic field and gyrofactor result in a smaller maximum electron Lorentz factor.
As the magnetic field is $B = 3.8$~G, the required gyrofactor is $\xi\sim10^8$ {under the assumption of the Bohm limit
acceleration}.
Therefore, our results imply not only the existence of a high magnetic field in the core, but also the very low particle acceleration efficiency in the core.

\begin{figure}[ht!]
  \centering
  \includegraphics[scale=0.45]{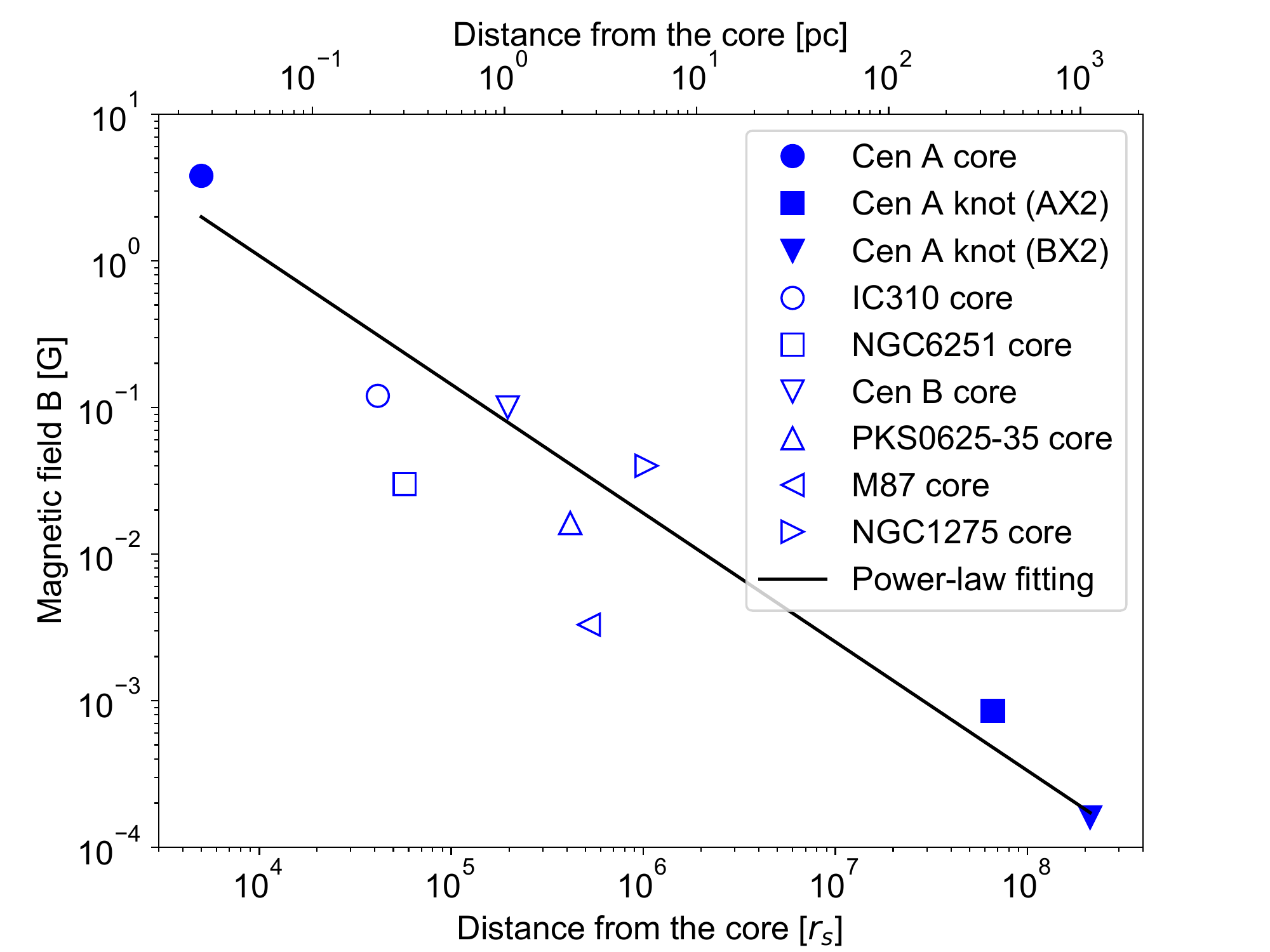}
  \caption{Magnetic fields obtained from the SED fits for various FR~I AGNs as functions of the distance from the core. The blue filled circle, square, and triangle represent the magnetic field strength of the Cen~A core, AX2, and BX2, respectively. The blue open marks indicate the magnetic field strength of the core of the other FR~I AGNs. The distances of AX2 and BX2 are directly measured using the $Chandra$ analysis \citep{2002ApJ...569...54K}. Except for these knots, the distance is calculated with the emission region size obtained by the SED model fitting as described in Section~\ref{sec:discussion}.
  The black solid line is the best-fit power-law function, $B (d) = B_0 (d/r_s)^{-\alpha}$, for these data.
  {For simplicity, the distance units for the other AGNs are also represented by the Schwarzschild radius of Cen A.}
  The obtained fitting parameters are the initial magnetic field of $B_0 = (1.5 \pm 0.2) \times 10^{4}$~G at the Schwarzschild radius $r_s$, and $\alpha=0.88 \pm 0.14$.}
  \label{fig:B_distance}
\end{figure}

This indicates that the emission region in the nucleus jet of Cen~A is located very close to the central core because the detailed general relativistic magnetohydrodynamics (MHD) simulation shows that the energy density of the magnetic field decays with the distance from the jet base (i.e., closer to the jet base the magnetic field is higher) \citep{2006MNRAS.368.1561M}.
In addition, the AGN jet can be collimated by the MHD process \citep{2001Sci...291...84M}, and the collimation profile of the M87 jet is recently constructed by the VLBI observations, which reports that the jet width follows $W_{\rm jet} \varpropto r^{0.56}$ up to $10^5$ times the Schwarzschild radius, and then, further downstream the jet transforms into a conical shape of $W_{\rm jet} \varpropto r^{1.0}$ \citep{2012ApJ...745L..28A, 2013ApJ...775...70H}.
Thus, the rather small source size of $R = 4.8 \times 10^{15}$~cm obtained from the SSC fitting for the core of Cen~A suggests that the emission region is located very close to the nucleus.
To verify that a higher magnetic field can be observed closer to the core, we plotted the obtained magnetic fields from the SED fits for various FR~I AGNs as functions of the distance from the core, as shown in Figure~\ref{fig:B_distance} {(no error is included)}.
It should be noted, that for the sake of simplicity, the distance of the emission region from the nucleus is defined as $d = R/\sin{\theta_{\rm op}}$, where $R$ is the source radius obtained from previous studies \citep{2003ApJ...597..166C, 2015MNRAS.450.4333D, 2017APh....89...14F, 2018Ap.....61..160B, 2018ApJ...860...74T, 2018CenB} and $\theta_{\rm op}$ is the opening angle of the jet. The VLBI observations revealed that the opening angle is collimated to the range of $2$--$8^\circ$ at a distance of 1~pc from the core; therefore, we set the angle as $5^\circ$ \citep{1999Natur.401..891J, 2006PASJ...58..211H}.
For the distances of the AX2 and BX2 knots, we used the directly measured values, as shown in Table~\ref{table:chandra_knots}.
As can be seen in Figure~\ref{fig:B_distance}, the magnetic field decays as a power-law function $B (d) = B_0 (d/r_s)^{-\alpha}$, where the Schwarzschild radius of Cen~A is $r_s = 2GM_{\rm BH}/c^2 \simeq 1.6 \times 10^{13}$~cm.
The obtained fitting parameters are the initial magnetic field of $B_0 = (1.5 \pm 0.2) \times 10^{4}$~G at $r_s$ and $\alpha=0.88 \pm 0.14$.
The conservation of particle number in a freely expanding conical jet requires the particle number density of $N (d) \varpropto d^{-2}$, which implies $B (d) \varpropto d^{-1}$ from equipartition \citep{2005ApJ...619...73H} and it is consistent with our results.
Furthermore, the core-shift analysis with the Very Long Baseline Array (VLBA) observations of blazars shows that the attenuation of the magnetic field at the vicinity of the core is described by $B (d) \varpropto d^{-1}$ appropriately \citep{2009MNRAS.400...26O}.
According to the theory of magnetically powered jet \citep{2007MNRAS.380...51K, 2010ASPC..427..207O}, the initial magnetic field strength for a black hole launched jet can be calculated by $B_0 \simeq 1.5 \times 10^{-5} L_{\rm rad}^{-1/2} r_s^{-1} \simeq 3.6 \times 10^3$~G, where $L_{\rm rad} \simeq 1.5 \times 10^{43} \ \rm erg \ s^{-1}$ is the total jet luminosity of Cen~A obtained from the SSC fitting, {which is about one-fourth of the fitted value. 
Although there is a discrepancy between the theory and fitted value which might be caused by the rough estimations such as the distance of the emission zones, these results support our suggestion that the SSC emission region of Cen A is located near the nucleus.
}

\section{Conclusions}\label{sec:conclusion}
In this paper, we proposed an SSC + EC/starlight model to explain the overall SED of Cen~A, including the hardening of the spectrum above the break energy of a few GeV.
This scenario assumes that the excess hard $\gamma$-ray emission results from the IC scattering of starlight photon fields by the electrons in the knots in the kpc-scale jet.
Remarkably, HESS observations recently reported that the TeV $\gamma$-ray emission is not point-like, but extended along the kpc-scale jet, which is consistent with our scenario \citep{2018TeVPA}. 
Considering the photon indices of the X-ray spectra analyzed using the $Chandra$ data in this work, we found that AX2 and BX2 knots can be appropriately related to the $\gamma$-ray spectrum (i.e., $\Gamma_{\rm X} < 2$).
Therefore, we fitted the excess GeV--TeV emission from the kpc-scale jet with AX2 and BX2.
As a result, the overall SED of Cen~A can be fitted appropriately by the SSC + EC/starlight model in both the AX2 and BX2 cases.
The one-zone SSC fitting for the core emission obviously requires a low maximum electron Lorentz factor of $\gamma_{\rm c} = 2.8 \times 10^3$ and large magnetic field of $B = 3.8$~G compared to that of other typical FR~I radio galaxies \citep{2003ApJ...597..166C, 2009ApJ...707...55A, 2018ApJ...860...74T, 2018Ap.....61..160B}.

Based on these results, we assume that the detection of the spectral hardening above the break energy of $\simeq 2.8$~GeV in Cen~A, unlike typical AGNs, is due to the following two reasons: (1) the host galaxy of Cen~A is strongly luminous and (2) the core $\gamma$-ray spectrum decays quickly above a few MeV due to low maximum electron Lorentz factor.
This is supported by the fact that the energy density of the starlight at the kpc-scale jet is sufficiently higher than those of other photons, such as CMB, synchrotron emission from the nucleus and the knot for IC scattering.
This suggests that a strong emission from the host galaxy is needed for producing bright GeV--TeV radiations.
In addition, we found that the magnetic field strength of relativistic jets approximately scales with the distance from the central black holes $d$ with $B (d) \propto d^{-0.88 \pm 0.14}$ as compared with other $\gamma$-ray detected radio galaxies.
Therefore, the high magnetic field, resulting in the low maximum electron Lorentz factor, indicates that the emission region of the inner jet is very close to the nucleus.
Based on these, we conclude that the IC radiation, peaking at $10$--$100$~GeV, becomes visible in Cen~A. 
This suggests that the EC radiation from the kpc-scale jets is hidden by the hard $\gamma$-ray spectrum in typical AGNs that have sufficiently large maximum electron Lorentz factors.
We plan to apply this scenario to detailed $\gamma$-ray analyses of various FR~I AGNs with increased sensitivity from the future Cherenkov Telescope Array observations \citep{2017arXiv170907997C}.

\acknowledgments 
Work by J.K. is supported in part by a project research hosted 
by Waseda Research Institute for Science and Engineering, 
also by Institute for Advanced Theoretical and Experimental 
Physics, Waseda university.
{We thank anonymous referees for careful reading our manuscript and for giving useful comments.}
\software
HEASoft (v6.19),
CIAO (v4.8)


\begin{thebibliography}{}
\bibitem[Abdo et al.(2009a)]{2009ApJ...699...31A} Abdo, A.~A., Ackermann, M., Ajello, M., et al.\ 2009, \apj, 699, 31.

\bibitem[Abdo et al.(2009b)]{2009ApJ...707...55A} Abdo, A.~A., Ackermann, M., Ajello, M., et al.\ 2009, \apj, 707, 55.

\bibitem[Abdo et al.(2010)]{2010ApJ...719.1433A} Abdo, A.~A., Ackermann, M., Ajello, M., et al.\ 2010, \apj, 719, 1433 

\bibitem[Abazajian et al.(2015)]{2015APh....63...55A} Abazajian, K.~N., Arnold, K., Austermann, J., et al.\ 2015, Astroparticle Physics, 63, 55 

\bibitem[Asada, \& Nakamura(2012)]{2012ApJ...745L..28A} Asada, K., \& Nakamura, M.\ 2012, \apj, 745, L28.

\bibitem[Baghmanyan et al.(2018)]{2018Ap.....61..160B} Baghmanyan, V., Tumanyan, M., Sahakyan, N., \& Vardanyan, Y.\ 2018, Astrophysics, 61, 160 

\bibitem[Balmaverde \& Capetti(2006)]{2006A&A...447...97B} Balmaverde, B., \& Capetti, A.\ 2006, \aap, 447, 97 

\bibitem[Becklin et al.(1971)]{1971ApJ...170L..15B} Becklin, E.~E., Frogel, J.~A., Kleinmann, D.~E., et al.\ 1971, \apjl, 170, L15

\bibitem[Beckmann et al.(2011)]{2011A&A...531A..70B} Beckmann, V., Jean, P., Lubi{\'n}ski, P., Soldi, S., \& Terrier, R.\ 2011, \aap, 531, A70 

\bibitem[Bednarek(2019)]{2019MNRAS.483.1003B} Bednarek, W.\ 2019, \mnras, 483, 1003.


\bibitem[Bennett et al.(2014)]{2014ApJ...794..135B} Bennett, C.~L., Larson, D., Weiland, J.~L., \& Hinshaw, G.\ 2014, \apj, 794, 135 

\bibitem[Brown et al.(2017)]{2017PhRvD..95f3018B} Brown, A.~M., B{\r{A}}`hm, C., Graham, J., et al.\ 2017, \prd, 95, 63018.

\bibitem[Cappellari et al.(2009)]{2009MNRAS.394..660C} Cappellari, M., Neumayer, N., Reunanen, J., et al.\ 2009, \mnras, 394, 660.

\bibitem[Cherenkov Telescope Array Consortium et al.(2017)]{2017arXiv170907997C} Cherenkov Telescope Array Consortium, T., :, Acharya, B.~S., et al.\ 2017, arXiv e-prints , arXiv:1709.07997.

\bibitem[Chiaberge et al.(2003)]{2003ApJ...597..166C} Chiaberge, M., Gilli, R., Capetti, A., \& Macchetto, F.~D.\ 2003, \apj, 597, 166 

\bibitem[de Jong et al.(2015)]{2015MNRAS.450.4333D} de Jong, S., Beckmann, V., Soldi, S., et al.\ 2015, \mnras, 450, 4333.

\bibitem[Fraija et al.(2017)]{2017APh....89...14F} Fraija, N., Marinelli, A., Galv{\'a}n-G{\'a}mez, A., et al.\ 2017, Astroparticle Physics, 89, 14.

\bibitem[Fraija et al.(submitted)]{2018CenB} Fraija, N., et al. \ submitted.

\bibitem[Fukazawa et al.(2011)]{2011ApJ...743..124F} Fukazawa, Y., Hiragi, K., Yamazaki, S., et al.\ 2011, \apj, 743, 124 

\bibitem[Ghisellini \& Tavecchio(2009)]{2009MNRAS.397..985G} Ghisellini, G., \& Tavecchio, F.\ 2009, \mnras, 397, 985 

\bibitem[Goodger et al.(2010)]{2010ApJ...708..675G} Goodger, J.~L., Hardcastle, M.~J., Croston, J.~H., et al.\ 2010, \apj, 708, 675.

\bibitem[Hada et al.(2013)]{2013ApJ...775...70H} Hada, K., Kino, M., Doi, A., et al.\ 2013, \apj, 775, 70.

\bibitem[Hardcastle et al.(2003)]{2003ApJ...593..169H} Hardcastle, M.~J., Worrall, D.~M., Kraft, R.~P., et al.\ 2003, \apj, 593, 169.

\bibitem[Hardcastle et al.(2007)]{2007ApJ...670L..81H} Hardcastle, M.~J., Kraft, R.~P., Sivakoff, G.~R., et al.\ 2007, \apj, 670, L81.

\bibitem[Hardcastle, \& Croston(2011)]{2011MNRAS.415..133H} Hardcastle, M.~J., \& Croston, J.~H.\ 2011, \mnras, 415, 133.

\bibitem[HESS~Collaboration et al.(2018)]{2018A&A...619A..71H} H.E.S.S.~Collaboration, Abdalla, H., Abramowski, A., et al.\ 2018, \aap, 619, A71 

\bibitem[Hirotani(2005)]{2005ApJ...619...73H} Hirotani, K.\ 2005, \apj, 619, 73.

\bibitem[Horiuchi et al.(2006)]{2006PASJ...58..211H} Horiuchi, S., Meier, D.~L., Preston, R.~A., \& Tingay, S.~J.\ 2006, \pasj, 58, 211 

\bibitem[Inoue \& Takahara(1996)]{1996ApJ...463..555I} Inoue, S., \& Takahara, F.\ 1996, \apj, 463, 555 

\bibitem[Inoue et al.(2013)]{2013ApJ...768..197I} Inoue, Y., Inoue, S., Kobayashi, M.~A.~R., et al.\ 2013, \apj, 768, 197 

\bibitem[Inoue \& Tanaka(2016)]{2016ApJ...828...13I} Inoue, Y., \& Tanaka, Y.~T.\ 2016, \apj, 828, 13 

\bibitem[Jarrett et al.(2003)]{2003AJ....125..525J} Jarrett, T.~H., Chester, T., Cutri, R., Schneider, S.~E., \& Huchra, J.~P.\ 2003, \aj, 125, 525 

\bibitem[Joshi et al.(2018)]{2018MNRAS.478L...1J} Joshi, J.~C., Miranda, L.~S., Razzaque, S., \& Yang, L.\ 2018, \mnras, 478, L1 

\bibitem[Junor et al.(1999)]{1999Natur.401..891J} Junor, W., Biretta, J.~A., \& Livio, M.\ 1999, \nat, 401, 891.

\bibitem[Kataoka et al.(1999)]{1999ApJ...514..138K} Kataoka, J., Mattox, J.~R., Quinn, J., et al.\ 1999, \apj, 514, 138.


\bibitem[Kellermann et al.(1997)]{1997ApJ...475L..93K} Kellermann, K.~I., Zensus, J.~A., \& Cohen, M.~H.\ 1997, \apjl, 475, L93 

\bibitem[Kraft et al.(2002)]{2002ApJ...569...54K} Kraft, R.~P., Forman, W.~R., Jones, C., et al.\ 2002, \apj, 569, 54.

\bibitem[Komissarov et al.(2007)]{2007MNRAS.380...51K} Komissarov, S.~S., Barkov, M.~V., Vlahakis, N., et al.\ 2007, \mnras, 380, 51.

\bibitem[Longair(2011)]{2011hea..book.....L} Longair, M.~S.\ 2011, High Energy Astrophysics.

\bibitem[Maraschi \& Tavecchio(2003)]{2003ApJ...593..667M} Maraschi, L., \& Tavecchio, F.\ 2003, \apj, 593, 667 

\bibitem[Markowitz et al.(2007)]{2007ApJ...665..209M} Markowitz, A., Takahashi, T., Watanabe, S., et al.\ 2007, \apj, 665, 209.

\bibitem[McKinney(2006)]{2006MNRAS.368.1561M} McKinney, J.~C.\ 2006, \mnras, 368, 1561 

\bibitem[Meier et al.(2001)]{2001Sci...291...84M} Meier, D.~L., Koide, S., \& Uchida, Y.\ 2001, Science, 291, 84 

\bibitem[O'Sullivan, \& Gabuzda(2009)]{2009MNRAS.400...26O} O'Sullivan, S.~P., \& Gabuzda, D.~C.\ 2009, \mnras, 400, 26.

\bibitem[O'Sullivan, \& Gabuzda(2010)]{2010ASPC..427..207O} O'Sullivan, S.~P., \& Gabuzda, D.~C.\ 2010, Accretion and Ejection in AGN: A Global View, 207.

\bibitem[Rybicki \& Lightman(1979)]{1979rpa..book.....R} Rybicki, G.~B., \& Lightman, A.~P.\ 1979, Radiative processes in astrophysics (New York: Wiley-Interscience), 393

\bibitem[Sahakyan et al.(2013)]{2013ApJ...770L...6S} Sahakyan, N., Yang, R., Aharonian, F.~A., \& Rieger, F.~M.\ 2013, \apjl, 770, L6 

\bibitem[Sanchez et al.(2018)]{2018TeVPA} Sanchez, D., Holler, M., Taylor, A. et al. for the HESS Collaboration 2018, talk during TeVPA2018 (Berlin)


\bibitem[Stawarz et al.(2003)]{2003ApJ...597..186S} Stawarz, {\L}., Sikora, M., \& Ostrowski, M.\ 2003, \apj, 597, 186 

\bibitem[Stawarz et al.(2006a)]{2006MNRAS.370..981S} Stawarz, {\L}., Aharonian, F., Kataoka, J., et al.\ 2006, \mnras, 370, 981 

\bibitem[Stawarz et al.(2006b)]{2006MNRAS.371.1705S} Stawarz, {\L}., Aharonian, F., Wagner, S., et al.\ 2006, \mnras, 371, 1705.

\bibitem[Steinle et al.(1998)]{1998A&A...330...97S} Steinle, H., Bennett, K., Bloemen, H., et al.\ 1998, \aap, 330, 97 

\bibitem[Steinle(2001)]{2001AIPC..587..353S} Steinle, H.\ 2001, Gamma 2001: Gamma-ray Astrophysics, 353.


\bibitem[Tanada et al.(2018)]{2018ApJ...860...74T} Tanada, K., Kataoka, J., Arimoto, M., et al.\ 2018, \apj, 860, 74.

\bibitem[Tingay et al.(1998)]{1998AJ....115..960T} Tingay, S.~J., Jauncey, D.~L., Reynolds, J.~E., et al.\ 1998, \aj, 115, 960.

\bibitem[Welch(1979)]{1979ApJ...228..419W} Welch, G.~A.\ 1979, \apj, 228, 419 



\end{thebibliography}
\end{document}